\documentclass[aps,prl,twocolumn,twoside,floatfix,superscriptaddress,preprintnumbers]{revtex4}

\usepackage{graphicx}
\usepackage{color}

\newcommand{\bea}{\begin{eqnarray}}
\newcommand{\beq}{\begin{equation}}
\newcommand{\eea}{\end{eqnarray}}
\newcommand{\eeq}{\end{equation}}

\begin{document}
\preprint{TUM-HEP-694/08}

\title{Low Energy Probes of CP Violation in a Flavor Blind MSSM}

\author{W.~Altmannshofer}
\affiliation{Physik-Department, Technische Universit\"at M\"unchen,
D-85748 Garching, Germany}

\author{A.J.~Buras}
\affiliation{Physik-Department, Technische Universit\"at M\"unchen,
D-85748 Garching, Germany}
\affiliation{TUM Institute for Advanced Study, Technische Universit\"at M\"unchen,
\\Arcisstr.~21, D-80333 M\"unchen, Germany}

\author{P.~Paradisi}
\affiliation{Physik-Department, Technische Universit\"at M\"unchen,
D-85748 Garching, Germany}

\begin{abstract}

We analyze the low energy implications of a {\it flavor blind} supersymmetric scenario
(where the CKM matrix is the only source of flavor violation) in the presence of new CP
violating but flavor conserving phases in the soft sector. We find that the best probes
of this rather restricted scenario are i) the electric dipole moments (EDMs) of the
electron ($d_e$) and the neutron ($d_n$) and ii) flavor changing and CP violating processes
in $B$ systems, like the CP asymmetries in $b\to s\gamma$ and $B\to\phi(\eta^{\prime})K_{S}$,
i.e. $A_{CP}(b\to s\gamma)$ and $S_{\phi(\eta^{\prime})K_{S}}$, respectively.
The non-standard values for $S_{\phi(\eta^{\prime})K_{S}}$, measured at the $B$~factories,
can find a natural explanation within our scenario and this would unambiguously imply i)
positive and often large (non-standard) values for $A_{CP}(b\to s\gamma)$ and ii) a lower bound
for the electron and neutron EDMs at the level of $d_{e,n}\gtrsim 10^{-28}\,e\,$cm.
Moreover, we predict positive New Physics (NP) contributions to $\epsilon_K$ which could be
welcomed in view of the recently lowered Standard Model value for $\epsilon_K$.
Interestingly, an explanation for the non-standard values for $S_{\phi(\eta^{\prime})K_{S}}$
can also naturally lead to an explanation for the anomaly of the muon anomalous magnetic moment.
Finally, we outline the role and the interplay of the direct NP searches at the LHC with the
indirect searches performed by low energy flavor physics observables.

\end{abstract}

\maketitle

\section{Introduction}\label{sec:intro}

In the last years, the two $B$ factories have established that $B_d$ flavor and CP violating
processes are well described by the Standard Model (SM) theory up to an accuracy of the
$(10-20)\%$ level. Unfortunately, irreducible hadronic uncertainties and the overall good
agreement of flavor data with the SM predictions still prevent any conclusive evidence of
NP effects in the quark sector.

This immediately implies a tension between the solution of the hierarchy problem and the 
explanation of the Flavor Physics data.

An elegant way to simultaneously solve the above problems is provided by the Minimal Flavor
Violation (MFV) hypothesis \cite{MFV,MFV_gen}, where flavor and CP violation are still entirely 
described by the CKM matrix.

This framework appears at first sight compatible with all the existing data.
On the other hand, a closer look at several CP violating observables indicates
that the CKM phase might not be sufficient to describe simultaneously CP violation 
in $K$, $B_d$ and $B_s$ decays. In particular:

i) Modes dominated by Penguin diagrams, such as $B\to (\phi, \eta^{\prime}, \pi^{0}, \omega,
K_{S}K_{S})K_{S}$ that, similarly to the golden mode $B\to \psi K_S$, allow the
determination of $\sin 2\beta$, result in $\sin 2\beta$ significantly lower than
$(\sin 2\beta)_{\psi K_S}= 0.680\pm 0.025$~\cite{hfag} from $B\to \psi K_S$. For the theoretical
cleanest modes it is experimentally found that $(\sin 2\beta)_{\phi K_S}= 0.39\pm 0.17$
and $(\sin 2\beta)_{\eta^{\prime} K_S}= 0.61\pm 0.07$~\cite{hfag}.

ii) With the decreased value of the non-perturbative parameter $\hat{B}_{K}$ from lattice
simulations \cite{hatBK} and the inclusion of additional negative contributions to $\epsilon_K$
that were neglected in the past \cite{BG}, CP violation in the $B_d-\overline{B}_d$ system,
represented by $(\sin 2\beta)_{\psi K_S}$, appears insufficient to describe the experimental
value of $\epsilon_K$ within the SM if the $\Delta M_d/\Delta M_s$ constraint is taken into
account~\cite{BG}.
Alternatively, simultaneous description of $\epsilon_K$ and $\Delta M_d/\Delta M_s$ within
the SM requires $\sin 2\beta= 0.88\pm 0.11$~\cite{SL}~\cite{BG}, significantly larger than
the measured $(\sin 2\beta)_{\psi K_S}$.

iii) There are some hints for the very clean asymmetry $S_{\psi\phi}$ to be significantly 
larger than the SM value $S_{\psi\phi}\approx 0.04$ \cite{first_evidence}.

iv) Finally, there is the muon anomalous magnetic moment anomaly. Most recent analyses converge
towards a $3\sigma$ discrepancy in the $10^{-9}$ range~\cite{g_2_th}:
$\Delta a_{\mu}\!=\!a_{\mu}^{\rm exp}\!-\!a_{\mu}^{\rm SM}\approx(3\pm 1)\times 10^{-9}$
where $a_{\mu}\!=\!(g-2)_{\mu}/2$.
Despite substantial progress both on the experimental~\cite{g_2_exp} and on the theoretical sides,
the situation is not completely clear yet. However, the possibility that the present discrepancy
may arise from errors in the determination of the hadronic leading-order contribution to
$\Delta a_{\mu}$ seems to be unlikely, as recently stressed in Ref.~\cite{passera_mh}.

There are also other interesting tensions observed in the data, as the rather large difference
in the direct CP asymmetries $A_{CP}(B^{-}\to K^{-}\pi^{0})$ and $A_{CP}(\overline{B}^{0}\to K^{-}\pi^{+})$
and certain puzzles in $B\to\pi K$ decays. However these tensions could also be due to our
insufficient understanding of hadronic effects rather than NP and we postpone their discussion
for the future.

The problems i)-iii) can easily be solved in any non-MFV framework like general MSSM~\cite{susy}
or Little Higgs models with T-parity~\cite{lht} (the MSSM also provides a natural explanation
for the problem iv)).
However, the large number of parameters in these extensions of the SM, does not allow for clear-cut 
conclusions.


\begin{table*}
\begin{tabular}{|l|l|l|l|}
\hline
Observable & SM Theory & Exp. Current & Exp. Future \\
\hline\hline
$S_{\phi K_S}$ & $\sin2\beta+0.02 \pm 0.01$~\cite{BDK_CERN} & $0.39 \pm 0.17$~\cite{hfag} & 
$(2-3)\%$~\cite{superb} \\
\hline
$S_{\eta^{\prime} K_S}$ & $\sin2\beta+0.01 \pm 0.01$~\cite{BDK_CERN} & $0.61 \pm 0.07$~\cite{hfag} & 
$(1-2)\%$~\cite{superb} \\
\hline
$A_{CP}(b\to s\gamma)$ & $\left(-0.44 ^{+0.14}_{-0.24}\right)\%$~\cite{hurth}& $\left(-0.4 \pm 3.6\right)\%$~\cite{hfag} & $(0.4-0.5)\%$~\cite{superb} \\
\hline
$|d_e|~~[e\,$cm] & $\approx 10^{-38}$~\cite{pospelov}& $< 1.6 \times 10^{-27}$ \cite{expedme} &
~$\approx 10^{-31}$~\cite{pospelov} \\
\hline
$|d_n|~~[e\,$cm] & $\approx 10^{-32}$~\cite{pospelov} & $< 2.9 \times 10^{-26}$ \cite{expedm} &
~$\approx 10^{-28}$~\cite{pospelov} \\
\hline
\end{tabular}
\caption{SM predictions and current/expected experimental sensitivities for the most relevant
observables for our analysis.}
\label{tab:observables}
\end{table*}


In this context, the question we intend to address in this paper is whether it is possible
to solve all these problems within a much more specific framework than a general MSSM,
namely a {\it flavor blind} supersymmetric scenario. In this framework, the CKM matrix remains
to be the only source of flavor violation but new CP violating, flavor conserving phases
are present in the soft sector.

In the present paper, we summarize the main results of our study. A more detailed presentation
will appear in~\cite{ABP2}.

One would naively expect that by far the best probes for CP violation within a {\it flavor blind}
MSSM (FBMSSM) are CP violating but flavor conserving observables, as the EDMs. As we will see,
this is not always the case and still large CP violating effects in flavor physics can
occur at an experimentally visible level. In particular, a large room for CP violating
asymmetries in $B$ systems, like $A_{CP}(b\to s\gamma)$ and $S_{\phi(\eta^{\prime})K_{S}}$,
is still possible, while $S_{\psi\phi}$ appears to remain small.

As we will see below, this framework is rather restrictive, when the data on 
${\rm BR}(b\to s\gamma)$, ${\rm BR}(B\to X_s\ell^+\ell^-)$, $\Delta M_{s,d}$, $\epsilon_K$,
$S_{\psi K_S}$ and $S_{\phi K_S}$ are simultaneously taken into account.
Moreover, the presence of flavor conserving but CP violating new phases implies definite 
correlations between NP effects in the observables described above and striking results 
for the electric dipole moments of the neutron and the electron and the direct CP asymmetry
$A_{CP}(b\to s\gamma)$ in the $b\to s\gamma$ decay. As the latter asymmetry is theoretically
very clean and very small in the SM, it constitutes similarly to $S_{\psi\phi}$ a very powerful
tool to search for new sources of CP violation.

Interestingly, this framework links the explanation of the suppression of $S_{\phi K_{S}}$
and $S_{\eta^{\prime} K_{S}}$ relative to $S_{\psi K_{S}}$ with the enhancement of
$A_{CP}(b\to s\gamma)$ over its SM value.

\section{Flavor Blind MSSM}\label{sec:FBMSSM}

The SM sources of CP violation are the QCD theta term $\overline{\theta}$
and the unique physical phase contained in the CKM matrix.
Natural (order one) values for $\overline{\theta}$ are phenomenologically
excluded since they would lead to unacceptably large contributions to the
neutron EDM.
Thus, a Peccei-Quinn symmetry \cite{peccei} is commonly assumed making $\overline{\theta}$
dynamically suppressed. In this way, the hadronic EDMs can be generated only by the
CP violating phase of the CKM and they turn out to be highly suppressed at the level
of $\sim 10^{-32}\,e\,$cm~\cite{pospelov}, well below the current and expected future 
experimental resolutions~\cite{expedm}.

On the other hand, the physical phase of the CKM successfully describes all the low
energy CP and flavor violating transitions so far observed in Nature, both in the
$K$ and $B_{d}$ systems except for possible tensions at the $2\sigma-3\sigma$ level
listed in i)-iii) that require further confirmations through improved data and theory
($\epsilon_K$).

Even though the SM certainly accounts for the bulk of CP violation in the $K$ and $B_{d}$
systems, we stress that it is still possible to expect spectacular NP phenomena to appear
in $B_s$ systems. In fact, on the one hand, the SM has not been experimentally tested in
$B_s$ systems with the same accuracy as in the $K$ and $B_d$ systems. On the other hand,
CP violating $b\to s$ transitions are predicted to be very small in the SM thus, any
experimental evidence of sizable CP violating effects in $B_s$ mixing would unambiguously
point towards a NP evidence.

Within a SUSY framework, CP violating sources may naturally appear after SUSY breaking
through i) flavor conserving $F$-terms (such as the $B\mu$ parameter in
the Higgs potential or the $A_I$ terms for trilinear scalar couplings) and ii) flavor
violating $D$-terms (such as the squark and slepton mass terms)~\cite{pospelov}.
It seems quite likely that the two categories i) and ii) of CP violation are controlled
by different physical mechanisms, thus, they can be distinguished and discussed independently.
In this paper we focus our attention on CP violating {\it flavor blind} phases, i.e. on the
category i). In such a case, it is always possible to choose a basis where only the $\mu$ and
$A_I$ parameters remain complex~\cite{pospelov}
and physics observables will depend only on the phases of the combinations
$M_i\mu$, $A_I\mu$ and $A_I^{\star}M_i$.

The new CP violating phases of the FBMSSM generally lead to too large effects for the electron
and neutron EDMs as they are induced already at the one loop level through the virtual exchange
of gauginos and sfermions of the 1st and 2nd generations.

There exist two natural ways to solve the above problem: i) to decouple the 1st and 2nd generation 
sfermions, as in effective SUSY models ~\cite{kaplan}, ii) to promote the $A_{I}$ terms as the main
source of CP violation and to assume they have a hierarchical structure, i.e. $A_t \gg A_c,A_u$,
$A_b\gg A_s,A_d$ and $A_{\tau} \gg A_{\mu},A_e$.

In the following, we analyze the latter situation postponing the study of the most general
case to our future works.

The flavor conserving phases of the FBMSSM are transmitted to the low energy observables by
means of the Feynmann diagrams shown in Fig.~\ref{c78}. On the left of Fig.~\ref{c78},
we report the dominant SUSY contribution for $S_{\phi K_{S}}$, $S_{\eta^{\prime} K_{S}}$
and $A_{CP}(b\to s\gamma)$ while, on the right, we show the two loop Barr-Zee type
diagram generating the EDM for quarks and leptons.

As it is already evident from these diagrams, the SUSY contributions of both $S_{\phi K_{S}}$,
$S_{\eta^{\prime} K_{S}}$, $A_{CP}(b\to s\gamma)$ and the EDMs are generated by the same CP
violating invariant $A_{t}\mu$.

\section{CP violation at the low energy}\label{sec:CP}

In what follows, we discuss the relevant observables for our study. The SM predictions,
as well as the current/expected experimental sensitivities of these observables are listed
in Table~\ref{tab:observables}.

\begin{figure}[t]
\centering
\includegraphics[scale=0.6]{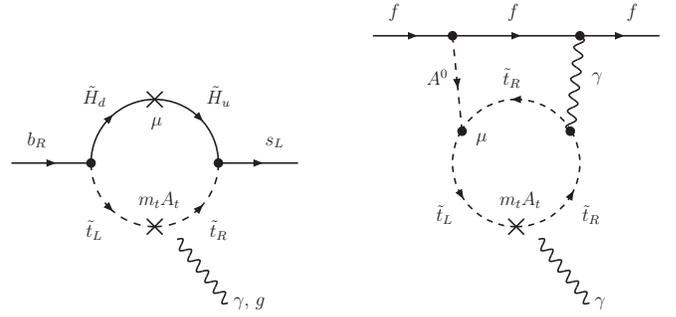}
\caption{
Left: Feynmann diagrams generating the dominant SUSY contributions to $S_{\phi K_{S}}$,
$S_{\eta^{\prime}K_{S}}$ and $A_{CP}(b\!\to\!s\gamma)$. Right: two loop Barr-Zee type
diagram generating an EDM for quarks ($f=q$) and leptons ($f=\ell$).}
\label{c78}
\end{figure}

{\bf 1.} As we have discussed above, one loop induced effects to the electron and neutron EDMs
can be always suppressed by assuming heavy 1st and 2nd generation sfermions and/or hierarchical
$A_{I}$ terms. However, even in these cases, additional contributions to the EDMs stemming from
two loop diagrams involving only the third sfermion generations are unavoidable and typically
large~\cite{pilaftsis} (see Fig.~\ref{c78} on the right for the stop contribution).
In particular, the expression for these two loop effects reads~\cite{pilaftsis}
\begin{equation}
{d_f \over e } = Q_f {3 \alpha_{\rm em} \over 32 \pi^2}
{R_f  m_f \over m_A^2} \sum_{q=t,b} \xi_q Q_q^2 F
\left({m_{\tilde{q}_1}^2\over m_A^2}, {m_{\tilde{q}_2}^2\over m_A^2 }\right)~,
\label{edmpil}
\end{equation}
\begin{equation}
\xi_t = \frac{m^{2}_{t}}{v^{2} s^{2}_{\beta}}\,
\frac{2{\rm Im}(\mu A_{t})}{m^{2}_{\tilde t_1}\!-\!m^{2}_{\tilde t_2}}~, ~~~
\xi_b = \frac{m^{2}_{b}}{v^{2} c^{2}_{\beta}}\,
\frac{2{\rm Im}(\mu A_{b})}{m^{2}_{\tilde b_1}\!-\!m^{2}_{\tilde b_2}}\,,
\end{equation}
where $F(x,y)$ is a two-loop function, $R_{t,b}=\cot\beta\,,\tan\beta$, $c_{\beta}=\cos\beta$, $s_{\beta}=\sin\beta$, $m_{\tilde{t}_{1,2}}$ ($m_{\tilde{b}_{1,2}}$) are the stop (sbottom)
masses and $m_A$ is the pseudoscalar Higgs mass.
The EDMs, as generated by means of Eq.~\ref{edmpil}, turn out to be large, provided
the heavy Higgs and/or 3rd generation squarks are not too heavy.

%


{\bf 2.} A very sensitive observable to CP violating effects is represented by the direct
CP violation in $b\to s\gamma$, i.e. $A_{CP}(b\to s\gamma)$ \cite{soares}. If NP effects
are present, the following expression for $A_{CP}(b\to s\gamma)$ holds~\cite{KN,hurth}
\bea
 A_{CP}(b\to s\gamma) &\equiv& 
\frac{\Gamma(B \to X_{\bar{s}} \gamma) - \Gamma(\overline{B} \to X_s \gamma)}{\Gamma(B \to X_{\bar{s}}\gamma) + \Gamma(\overline{B} \to X_s\gamma)} \nonumber \\
&\simeq&
{-1 \over | C_7 |^2 }~\!\bigg(\!1.23\,{\rm Im}[C_2 C_7^*]-9.52\,{\rm Im}[C_8 C_7^*]\nonumber \\
&+& 0.10 ~{\rm Im}\,[C_2 C_8^*]\bigg) - 0.5~({\rm in}~\%)~,
\label{acp_bsgamma}
\eea
where $C_i\!=\!C^{\rm SM}_{i}(m_b)+C^{\rm NP}_{i}(m_b)$, with $C^{\rm SM}_{i}$ real.
In our framework, the dominant SUSY contributions to $C_{7,8}$ arise from the one-loop
charged Higgs and chargino-squark amplitudes:
$C^{\rm NP}_{7,8}=C^{H^\pm}_{7,8}+C^{\tilde{\chi}^\pm}_{7,8}$.

In Fig.~\ref{c78}, on the left, we report the Feynmann diagrams for the dominant chargino-squark
contributions in the so-called mass insertion approximation. As we can see, it turns out that
$C^{\tilde{\chi}^\pm}_{7,8}\sim\mu A_{t}$ thus $C^{\tilde{\chi}^\pm}_{7,8}$ develop an imaginary
component when ${\rm Im}\,\mu A_{t}\neq 0$.
The corresponding approximate expressions for the charged Higgs and chargino contributions
to $C_7$ and $C_8$ read
\bea
C_{7,8}^{H^\pm}\simeq f^1_{7,8}(x_t)~,\qquad
C^{\tilde{\chi}^\pm}_{7,8}\simeq
\frac{m_t^2}{m_{\tilde{q}}^2}\bigg(\frac{A_t\mu}{m_{\tilde{q}}^2}\bigg)t_{\beta}f_{7,8}^{2}(x_\mu)~,
\label{C_7}
\eea
%
%
%
where $x_t=m_t^2/M_{H^\pm}^2$, $x_{\mu}=|\mu|^2/m^{2}_{\tilde{q}}$, $f^1_7(1)=-7/36$,
$f^1_8(1)=-1/6$, $f^2_7(1)=-5/72$, $f^2_8(1)=-1/24$, $t_{\beta}=\tan\beta$ and
$m_{\tilde q}^2$ is an average stop mass.

In our numerical analysis, we have also included higher order threshold corrections
stemming from $\tan\beta$ enhanced non-holomorphic Yukawa interactions that are not
explicitly shown in Eq.~\ref{C_7}.

These threshold corrections allow $C^{H^\pm}_{7,8}$, that are purely real at the leading
order, to develop an imaginary component and provide additional CP violating contributions
to $C^{\tilde{\chi}^\pm}_{7,8}$, that are complex already at the leading order.


{\bf 3.} The time-dependent CP asymmetries in the decays of neutral $B$ mesons into final CP
eigenstates $f$ are described by ${\cal A}_f(t)=S_f\sin(\Delta M t)-C_f\cos(\Delta M t)$,
where, within the SM, it is predicted that the $|S_f|$ and $C_f$ parameters relative to all
the transitions $\bar b\to\bar q^\prime q^\prime\bar s$ ($q^\prime=c,s,d,u$) are the same.
In particular, the SM predicts that $-\eta_f S_f\simeq\sin2\beta$ and $C_f\simeq0$ where 
$\eta_f=\pm1$ is the CP  eigenvalue for the final state $f$, and $\beta\equiv\arg\left[-(V_{cd}V_{cb}^*)/(V_{td}V_{tb}^*)\right]$.
In order to define the CP asymmetries in $B\to f$ decays, one introduces the complex quantity
$\lambda_f=e^{-i\phi_B}(\overline{A}_f/A_f)$ where $\phi_B$ is the phase of the $B^0-\overline{B}^0$
mixing amplitude, $A_f$ ($\overline{A}_f$) is the decay amplitude for $B^0(\overline{B}^0)\to f$ 
and finally
\beq
S_f=\frac{2{\rm Im}(\lambda_f)}{1+|\lambda_f|^2}~,\ \ \
C_f=\frac{1-|\lambda_f|^2}{1+|\lambda_f|^2}~.
\eeq
The decay amplitudes can be calculated from the effective Hamiltonian relevant for
$\Delta B=\pm1$ decays~\cite{BBL} through $A_f=\langle f|{\cal H}_{\rm eff}|B^0\rangle~$
and $\overline{A}_f=\langle f|{\cal H}_{\rm eff}|\overline{B}^0\rangle$, where the
Wilson coefficients depend on the electroweak theory while the matrix elements $\langle f|O_i|B^0(\overline{B}^0)\rangle$ can be evaluated, for instance, by means of QCD factorization.

New Physics effects can contribute i) to the $B^0-\overline{B}^0$ mixing amplitude 
or ii) to the decay amplitudes $\bar b\to\bar qq\bar s$ ($q=s,d,u$)~\cite{grossman}
(we assume the tree level transition $\bar b\to\bar cc\bar s$,
and thus also the related asymmetry, to be not significantly affected by the NP).
In presence of NP contributions, we can write the decay amplitudes in the following way
\beq\label{defbfu}
A_f = A_f^c
\left[1+a_f^ue^{i\gamma}+\sum_i\left(b_{fi}^c+b_{fi}^ue^{i\gamma}\right)C_i^{\rm NP\,*}(m_W)\right].
\eeq
%
The $a_f^u$ parameters have been evaluated in the QCD factorization approach at the leading
order and to zeroth order in $\Lambda/m_b$ in Ref.~\cite{buchalla}.
Within the SM, it turns out that $S_{\phi K_S}\!\simeq\!S_{\eta^{\prime}K_{S}}\!\simeq\!
S_{\psi K_S} \simeq 0.68$ (see Table~\ref{tab:observables}).
The $a_f^u$ term provide only a negligible effect to $B\to\psi K_S$, thus
$\lambda_{\psi K_S}^{\rm SM}=-e^{-2i\beta}$.

Within a FBMSSM scenario, it turns out that the Wilson coefficient $C^{\rm NP}_{8}$ of the
chromomagnetic operator provides the dominant NP source for $S_{\phi(\eta^{\prime})K_S}$.
In particular, all the Wilson coefficients but the electromagnetic and chromomagnetic ones
are not sensitive to the new phases of the FBMSSM.
The relevant hadronic parameters entering the $C^{\rm NP}_{8}$ contribution to 
$S_{\phi(\eta^{\prime})K_S}$ are $b_{\phi K_{S}}^{c}= 1.4$ and $b_{\eta^{\prime}K_{S}}^{c}= 0.86$~\cite{buchalla}, thus, the NP effects in $S_{\phi K_S}$ are significantly larger than 
those in $S_{\eta^\prime K_S}$. Moreover, the departures from the SM expectations of both 
$S_{\phi K_S}$ and $S_{\eta^{\prime}K_S}$, due to NP contributions, are expected to be in the 
same direction.\\


{\bf 4.} Also $\epsilon_K$ is a NP sensitive observable of our scenario and it can receive NP
effects at the $15\%$ level compared to its SM prediction. Interestingly, the FBMSSM
scenario unambiguously predicts $|\epsilon_K|>|\epsilon_K^{\rm SM}|$~\cite{ABP2} and this helps
to achieve the agreement with data.
The same conclusion is no longer valid for the $B$ systems. However, it is predicted that
the ratio $\Delta M_d/\Delta M_s$, as well as $S_{\psi K_S}$, are SM-like to a very good
approximation~\cite{ABP2}.\\

\section{Numerical analysis}

In order to establish the allowed CP violating effects in $B$ physics observables, we impose
the full set of available theoretical and experimental constraints. In particular, we take
into account i) the constraints from direct search, ii) the electroweak precision tests
(as the $\rho$--parameter), iii) the requirements of color and electric charge conservation
and that the lightest SUSY particle (LSP) has to be neutral, iv) the constraints from the
EDM of the electron/neutron, v) all the constraints from flavor changing processes in $K$ and
$B$ physics.

In our scenario, the EDMs and the branching ratio of $b\to s\gamma$ constitute the two most severe
constraints. Concerning the latter, combining the SM prediction at the NNLO~\cite{Misiak} with the
experimental average for the branching ratio~\cite{hfag,belle,babar} we obtain
\beq
R_{bs\gamma} =
 \frac{ {\rm BR}(b\to s \gamma)^{\rm exp}}
{{\rm BR}(b\to s \gamma)^{\rm SM}}
= 1.13\pm 0.12~.
\label{bsgamma}
\eeq
In our numerical analysis we impose the above constraint at the $2\sigma$ C.L..

In our FBMSSM, we assume universal soft masses for different squark generations at
the electroweak scale~\footnote{Such a strong assumption gets somewhat relaxed in the
framework of the general MFV ansatz~\cite{MFV_gen} where the scalar soft masses receive 
corrections proportional to $c_i\,Y^\dagger Y$ (where $Y$ denotes a Yukawa
matrix and $c_i$ are unknown coefficients typically of order ${\mathcal O}(1)$).
These small departures from a complete flavor blindness of the soft terms generate
additional FCNC contributions by means of gluino and squark loops. However, these
last effects can be safely neglected if the unknown parameters $c_i$ are small
and/or if the gluino mass is significantly larger than the chargino/up-squark masses.
In this respect, the contributions to FCNC processes discussed in the present work
can be regarded as irreducible effects arising in MFV scenarios.}.

\begin{figure}
\includegraphics[scale=0.33]{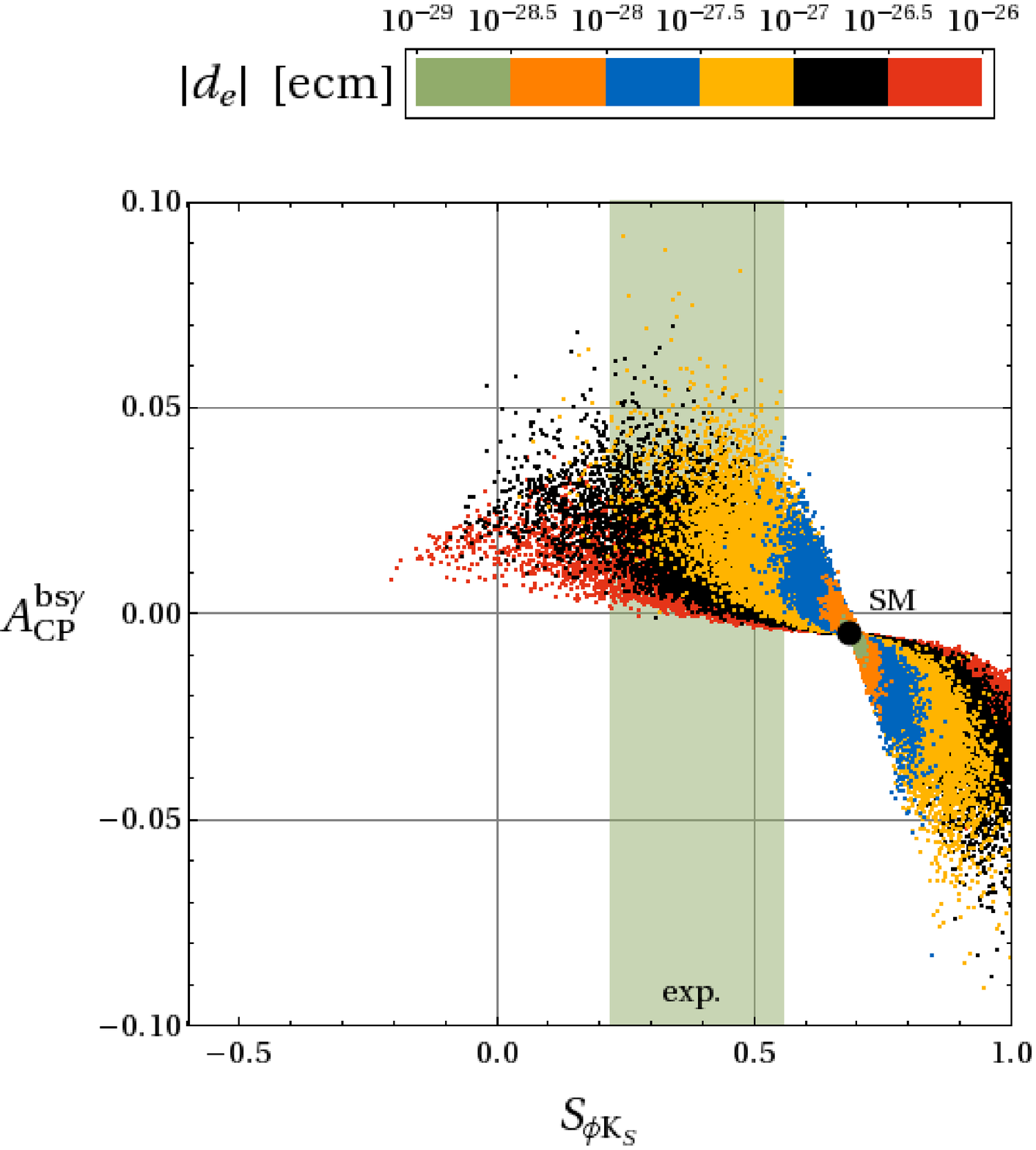} \\[16pt]
\includegraphics[scale=0.35]{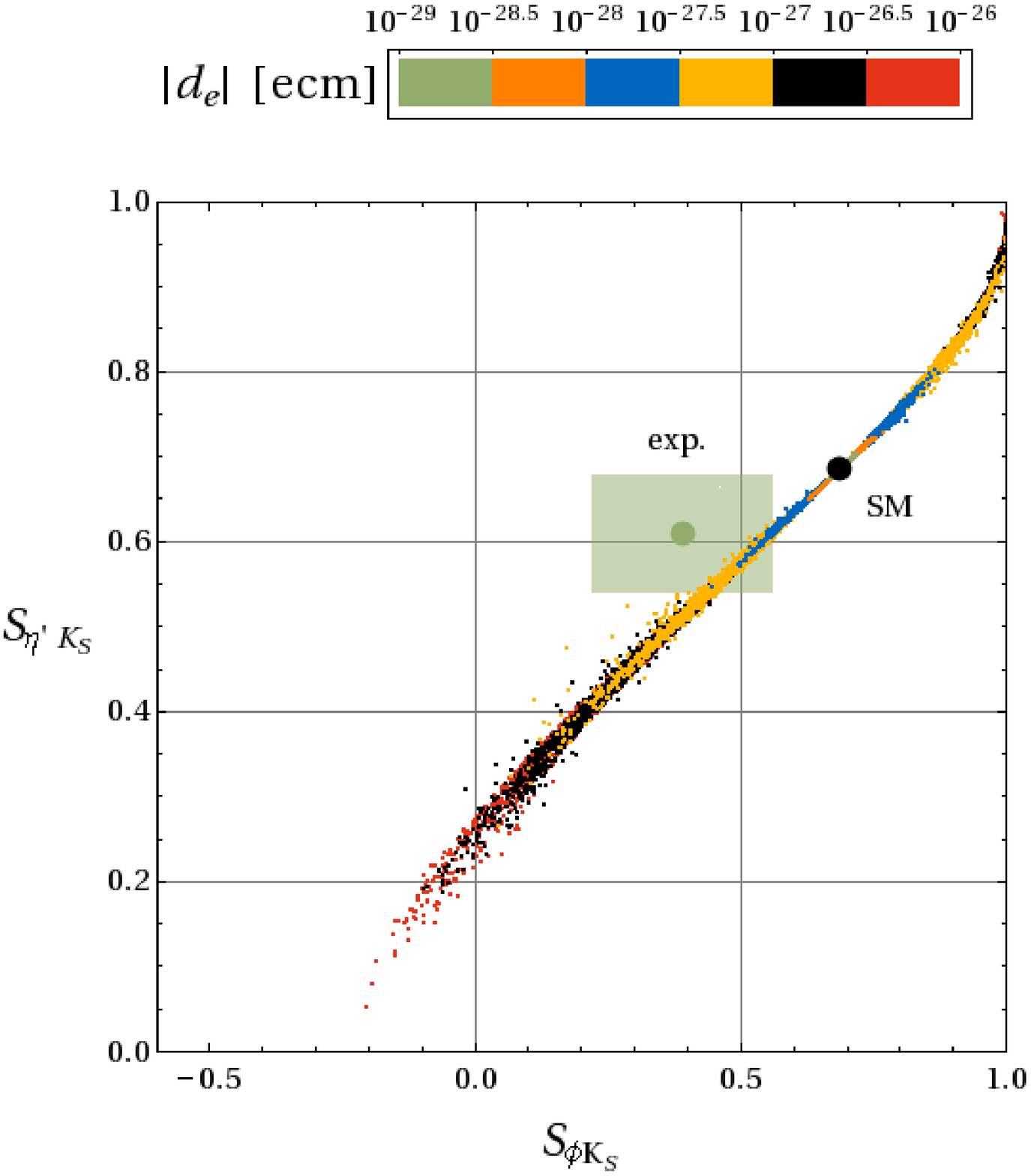}
\caption{
Upper: $A_{CP}(b\to s\gamma)$ vs $S_{\phi K_{S}}$.
Here, as well as in all the other plots, the grey band corresponds to the current
experimental bounds on $S_{\phi K_{S}}$ at the $68\%$ C.L.
Lower: $S_{\eta^{\prime}K_{S}}$ vs $S_{\phi K_{S}}$.
The SM value for $S_{\phi K_{S}}\simeq S_{\eta^{\prime}K_{S}}\simeq 0.68$ is also
indicated. The grey box corresponds to the current $68\%$ C.L.. experimental bounds
on $S_{\phi K_{S}}$ and $S_{\eta^{\prime}K_{S}}$.
In both plots, the attained values for the electron EDM $d_{e}$ are also shown.
}
\label{sphiks_cpbsg}
\end{figure}

After imposing the above constraints, we perform a scan over the relevant SUSY parameter space
\bea
&(\mu,M_{H^+}, m_{\tilde{t}_{L}}, m_{\tilde{t}_{R}}, m_{\tilde{g}})\leq 1~{\rm TeV}\,,&\nonumber \\
&|A_t|^{2}\leq 3 (m^{2}_{\tilde{t}_{L}}+m^{2}_{\tilde{t}_{R}})\,,& \nonumber \\
&3\leq t_{\beta}\leq 50\,,& \nonumber \\
&0\leq \phi_{\mu}+\phi_{A_t}\leq 2\pi\,.&
\label{scan}
\eea
In the upper plot of Fig.~\ref{sphiks_cpbsg}, we show $A_{CP}(b\to s\gamma)$ vs $S_{\phi K_{S}}$.
First, we emphasize that non-standard values for $S_{\phi K_{S}}$ are easily achieved in the
FBMSSM framework and the present $S_{\phi K_{S}}$ anomaly can find a natural explanation.
Moreover, the sign of $S_{\phi K_{S}}$ is correlated with the sign of $A_{CP}(b\to s\gamma)$.
In particular, in the preferred experimental region for $S_{\phi K_{S}}$, i.e. for
$S_{\phi K_{S}}< S_{\phi K_{S}}^{\rm SM}\approx 0.68$, it turns out that $A_{CP}(b\to s\gamma)$ is dominantly {\it positive} and has therefore opposite sign relative to the SM expectation.
Interestingly enough, in the region of the parameter space where $S_{\phi K_{S}} \approx 0.4$,
as suggested experimentally, $A_{CP}(b\to s\gamma)$ is typically predicted to depart significantly
from its SM prediction.

\begin{figure}
\includegraphics[scale=0.41]{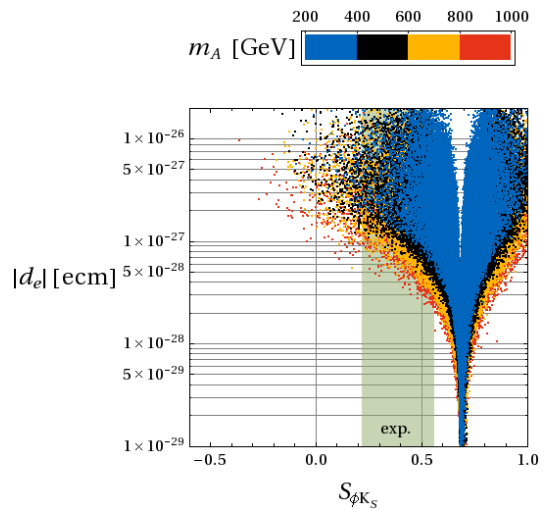}
\caption{
Electron EDM ($d_e$) vs $S_{\phi K_{S}}$.
The colored bands correspond to different values for the pseudoscalar Higgs mass $m_{A}$.
}
\label{edm_etaphi}
\end{figure}

In the lower plot of Fig.~\ref{sphiks_cpbsg}, it is shown the correlation between $S_{\phi K_{S}}$
and $S_{\eta^{\prime} K_{S}}$. As we can see, the NP effects in $S_{\phi K_{S}}$ are larger than
those in $S_{\eta^{\prime} K_{S}}$ in agreement with the pattern observed in the data.

In both plots of Fig.~\ref{sphiks_cpbsg}, the various colored bands show the attained values for
the electron EDM $d_e$.

We believe that a correlated analysis of the above asymmetries at a Super Flavor Factory
would represent a powerful tool to probe or to falsify the FBMSSM scenario.

In Fig.~\ref{edm_etaphi}, we show the prediction for the electron EDM $d_e$ vs $S_{\phi K_{S}}$
for different values of $m_{A}$. We note that $d_e$ is very sensitive to $m_A$, as the NP
contributions to $d_e$ decouple with the heaviest mass between $m_A$ and the stop masses
$m_{\tilde{t}_{1,2}}$ (see Eq.~\ref{edmpil}).
Fig.~\ref{edm_etaphi} shows that large (non-standard) effects in $S_{\phi K_{S}}$ unambiguously
imply large values for $d_e$.
In particular, in the experimentally interesting region where $S_{\phi K_{S}}\approx 0.4$,
we obtain the lower bound $d_e\geq 5\times 10^{-28}\,e\,$cm by means of the scan over
the SUSY parameter space of Eq.~\ref{scan}. Similar results are found for the neutron EDM
$d_n$, in which case, we find the lower bound $d_n\geq 8\times 10^{-28}\,e\,$cm.

However, we observe that while $S_{\phi K_{S}}$, $S_{\eta^{\prime}K_{S}}$ and $A_{CP}(b\to s\gamma)$
are not directly sensitive to $m_{A}$ (they feel $m_{A}$ mainly through the indirect
${\rm BR}(b\to s\gamma)$ constraint), in contrast, $d_{e,n}$ go to zero when $m_{A}$ decouples.
In particular, if we enlarge the allowed values for $m_{A}$ up to $m_{A}\!<\!3~\rm{TeV}$ while
varying all the other SUSY parameters in the same range as in Eq.~\ref{scan}, the requirement of
$S_{\phi K_{S}}\approx 0.4$ would imply the lower bound $d_e\geq (5\,,3\,,2\,,1\,,0.5)\times
10^{-28}\,e\,$cm for $m_{A}\leq(1\,,1.5\,,2\,,2.5\,,3)~\rm{TeV}$, respectively.

\begin{figure}
\includegraphics[scale=0.35]{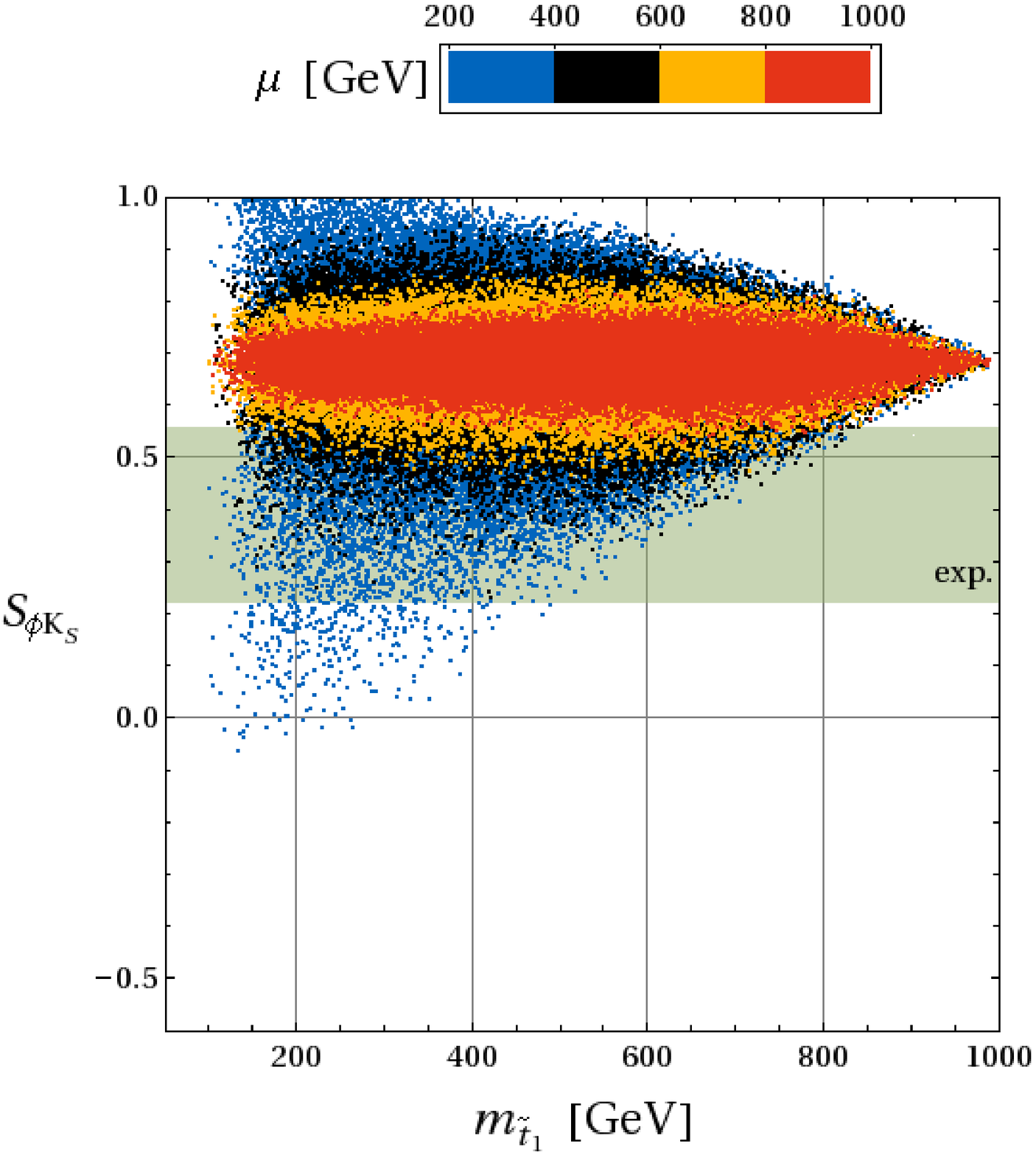}\\[16pt]
\includegraphics[scale=0.32]{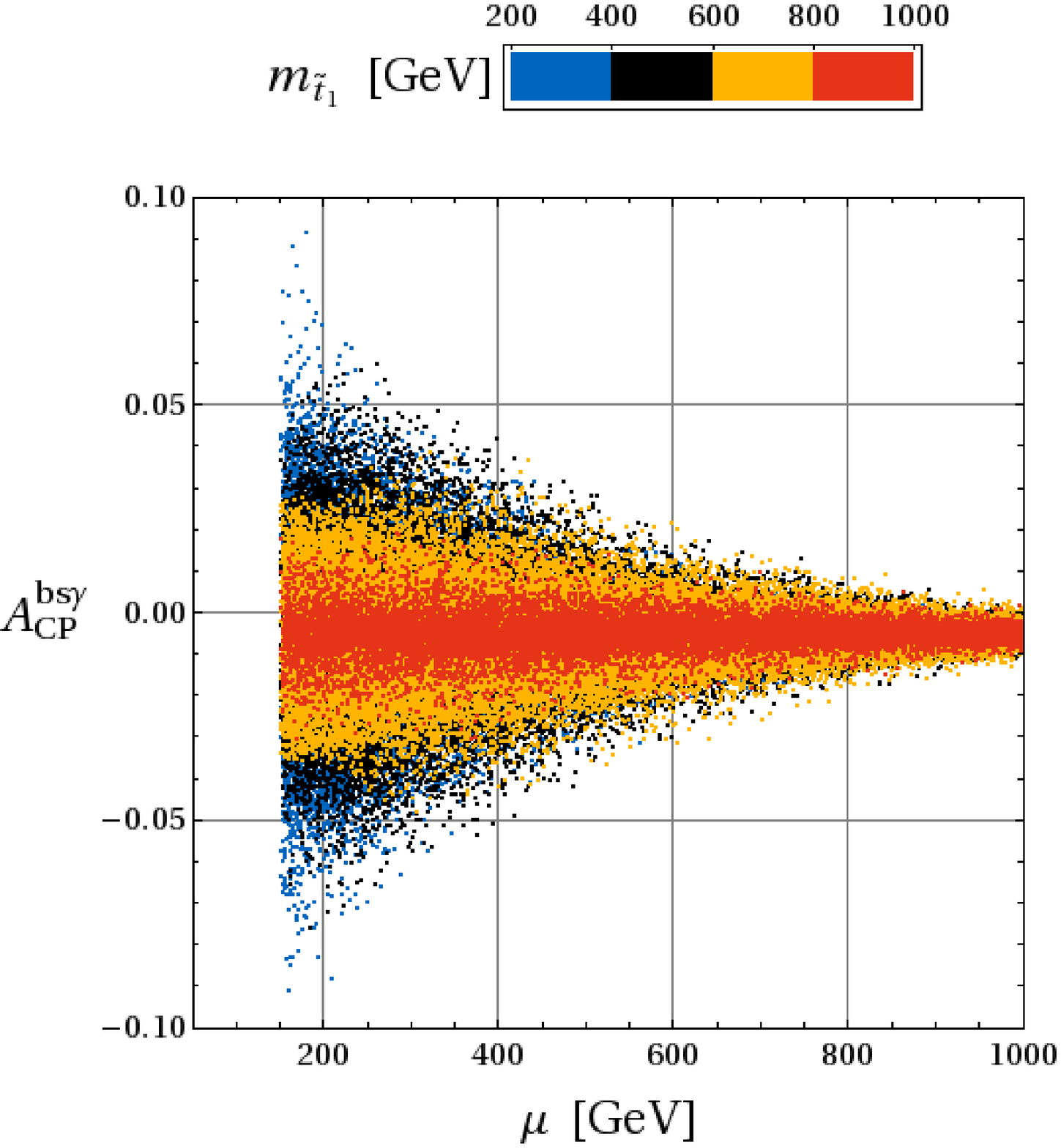}
\caption{
Upper: $S_{\phi K_{S}}$ vs $m_{\tilde{t}_{1}}$. Lower: $A_{CP}(b\to s\gamma)$ vs $\mu$.
The dependence of $S_{\phi K_{S}}$ ($A_{CP}(b\to s\gamma)$) on the $\mu$ parameter
(lightest stop mass $m_{\tilde{t}_{1}}$) is also shown through different colored bands.
}
\label{sphiks_mstop}
\end{figure}

In Fig.~\ref{sphiks_mstop}, upper plot, we show the dependence of $S_{\phi K_{S}}$ on the
lightest stop mass $m_{\tilde{t}_{1}}$ for different values of the $\mu$ parameter. We see
that large (non-standard) effects for $S_{\phi K_{S}}$ can be expected even for a SUSY
spectrum at the \rm{TeV} scale.

In Fig.~\ref{sphiks_mstop}, lower plot, we show $A_{CP}(b\to s\gamma)$ as a function of the
$\mu$ parameter for different values of the lightest stop mass $m_{\tilde{t}_{1}}$. We note
that, $A_{CP}(b\to s\gamma)$ can reach non-standard values $|A_{CP}(b\to s\gamma)|\!>\!2\%$
only if $\mu \lesssim 600-700$~GeV, well within the LHC reach.

\begin{figure}
\includegraphics[scale=0.32]{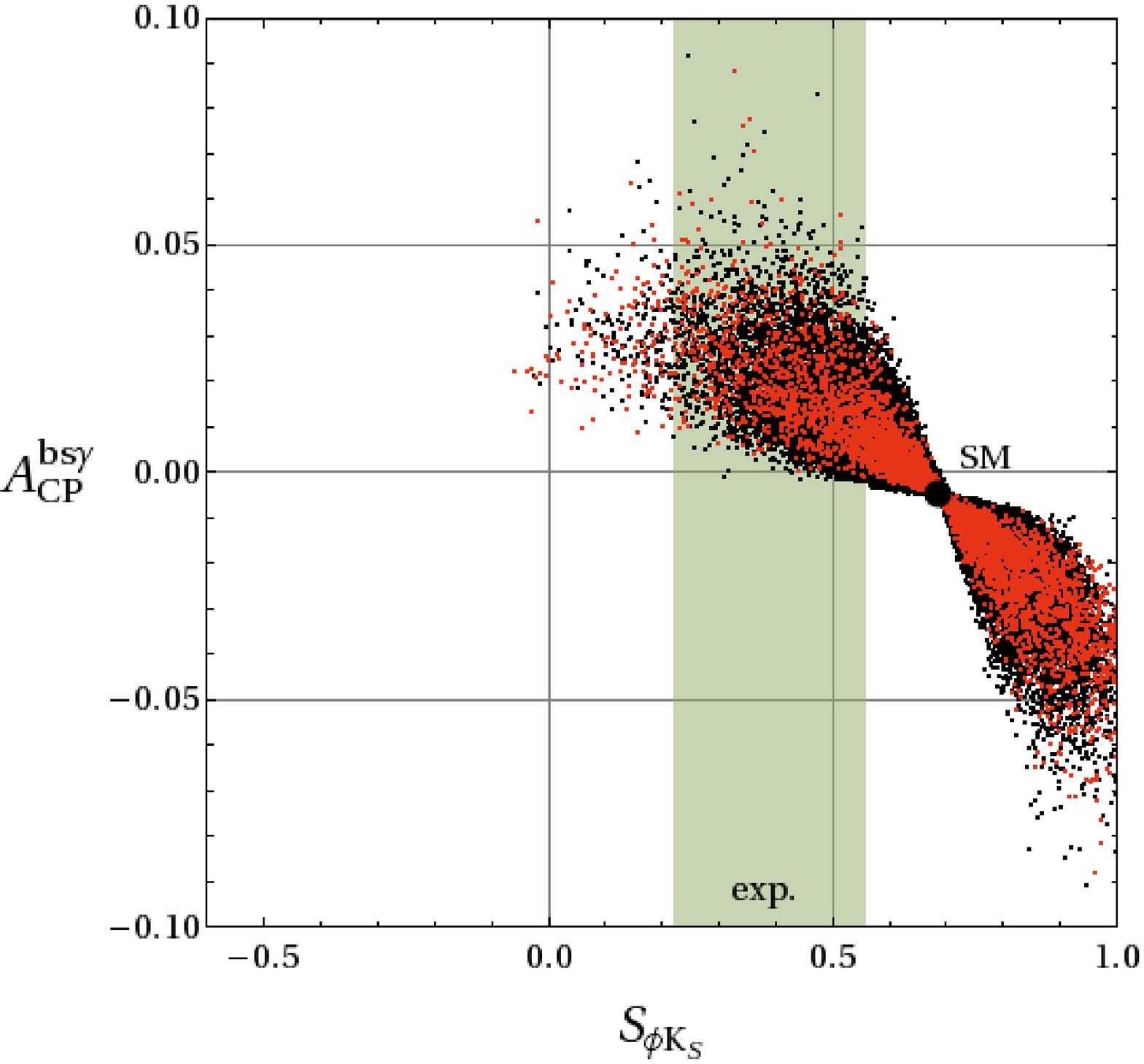}\\[16pt]
\includegraphics[scale=0.35]{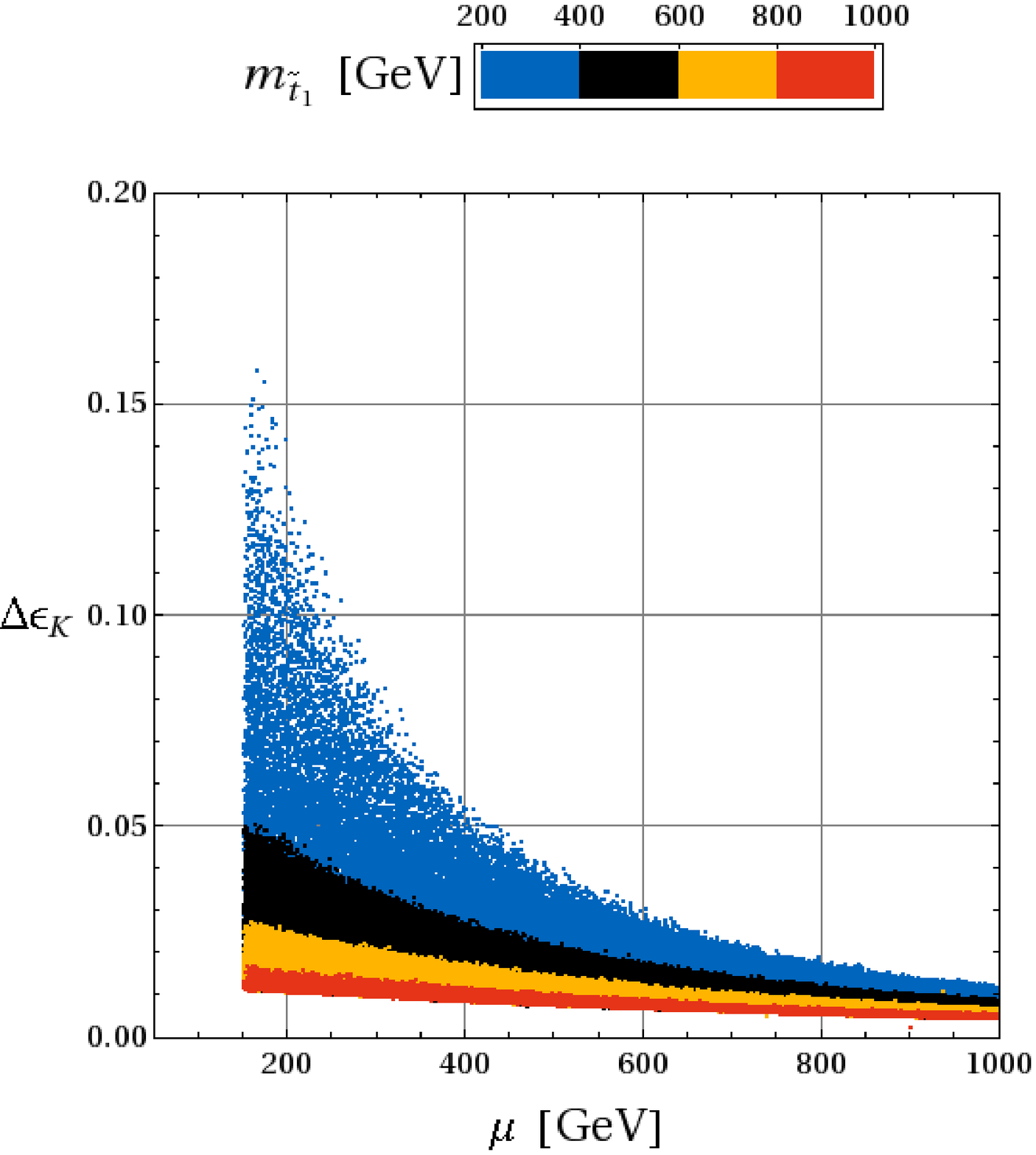}
\caption{
Upper: $A_{CP}(b\to s\gamma)$ vs $S_{\phi K_{S}}$. Red dots satisfy $\Delta\epsilon_K\!>\!5\%$.
Lower: $\Delta\epsilon_K$ vs $\mu$. The values attained by $\Delta\epsilon_K$ for different
lightest stop masses $m_{\tilde{t}_{1}}$ are also shown.}
\label{epsk}
\end{figure}

In Fig.~\ref{epsk}, upper plot, we show again $A_{CP}(b\to s\gamma)$ vs $S_{\phi K_{S}}$ selecting
the points (red dots) satisfying 
$\Delta\epsilon_K\!=\!\epsilon^{\rm{SUSY}}_{K}/\epsilon^{\rm{SM}}_{K}\!-1\!\!>\!5\%$.
As we can see, in the experimentally interesting region where $S_{\phi K_{S}}\approx 0.4$, there
are many points where also $\epsilon_K$ can receive sizable NP effects up to corrections
of order $\approx 15\%$ compared to the SM contributions (see the lower plot of Fig.~\ref{epsk}).
However, there is not a direct correlation between $S_{\phi K_{S}}$ and $\Delta\epsilon_K$, as the
latter is not sensitive to the new phases of the FBMSSM, in contrast to $S_{\phi K_{S}}$~\cite{ABP2}.

In the lower plot of Fig.~\ref{epsk}, we show $\Delta\epsilon_K$ as a function of the $\mu$ parameter.
Moreover, different colored bands show the dependence of $\Delta\epsilon_K$ on the lightest stop
mass $m_{\tilde{t}_{1}}$. As we can see, sizable NP effects for $\epsilon_K$ at the level of $\Delta\epsilon_K\approx 15\%$ can be still obtained for a light SUSY spectrum 
$(\mu,\,m_{\tilde{t}_{1}})\approx 200~\rm{GeV}$.

Finally, let us briefly discuss the $\Delta a_{\mu}$ anomaly within our scenario.
Even though $\Delta a_{\mu}$ does not require in itself any source of CP violation to
occur, it is anyway interesting to observe that large NP effects in $B\to\phi K_S$,
as required by the current experimental data, typically point towards sizable effects
also in $\Delta a_{\mu}$. To see this point explicitly, we report the expression for
the SUSY contributions to $\Delta a_{\mu}$~\cite{moroi} in the limit where all the SUSY 
masses are degenerate with a common mass $m_{\tilde \ell}$. In such a case one finds
\beq
\frac{a^{\rm MSSM}_\mu}{ 1 \times 10^{-9}}
\simeq 1.5\left(\frac{\tan\beta }{10} \right)
\left( \frac{300~\rm GeV}{m_{\tilde \ell}}\right)^2 \rm{sign}(\rm{Re}\,\mu)\,,
\label{eq:g_2}
\eeq
clearly showing that for SUSY masses of order few hundred $\rm{GeV}$ and moderate to large
$\tan\beta$ values, the anomaly $\Delta a_{\mu}\!=\!a_{\mu}^{\rm exp}\!-\!a_{\mu}^{\rm SM}
\approx(3\pm 1)\times 10^{-9}$ can be easily explained, provided $\rm{Re}\,\mu\!>\!0$.

On the other hand, within a FBMSSM scenario, the chromomagnetic operator $C^{\rm NP}_{8}$ provides
the dominant NP source for $S_{\phi(\eta^{\prime})K_S}$ and, in particular, it turns out that
${\rm Im}C^{\rm NP}_{8}\!\sim\!(m_t/m_{\tilde t})^{2}\,\tan\beta\,\sin(\phi_{A_t}+\phi_{\mu})$.
Under the mild assumption that $m_{\tilde \ell}$ is not much heavier than $m_{\tilde t}$
(as it is the case in most of the SUSY breaking mechanisms), and even assuming maximum CP violation,
i.e. $|\sin(\phi_{A_t}+\phi_{\mu})|\sim 1$, we find that $S_{\phi(\eta^{\prime})K_S}\sim 0.4$ naturally
leads to $a^{\rm MSSM}_\mu \!>\! 10^{-9}$, in accordance with the experimental data.

\section{Conclusions}

In the present paper we have investigated the low energy implications of a {\it flavor blind}
supersymmetric scenario (where the CKM matrix is the only source of flavor violation) in the
presence of new CP violating but flavor conserving phases in the soft sector.

In particular, we have analyzed their impact on flavor conserving but CP violating transitions
like EDMs and also on a large number of flavor and CP violating observables taking into account
all the existing theoretical and phenomenological constraints.

The NP scenario in question is characterized by a number of NP parameters that is much smaller
than encountered in general SUSY models, LHT models and models with a warped extra dimension.
This implies striking correlations between various observables that can confirm or exclude this
scenario in coming years.

We find that $S_{\phi K_{S}}$ and $S_{\eta^{\prime} K_{S}}$ can both differ from $S_{\psi K_{S}}$
and can be larger or smaller than $S_{\psi K_{S}}$ with the effect being typically by a factor of
$1.5$ larger in $S_{\phi K_{S}}$ in agreement with the pattern observed in the data.
Most interestingly, we find that the desire of reproducing the observed low values of $S_{\phi K_{S}}$
and $S_{\eta^{\prime} K_{S}}$ implies uniquely:

i) Lower bounds on the electron and neutron EDMs $d_{e,n} \gtrsim 10^{-28}\,e\,$cm.

ii) Positive and sizable (non-standard) $A_{CP}(b\to s\gamma)$ asymmetry in the ballpark
of $1\%-5\%$, that is opposite sign to the SM one.

iii) The NP effects in $S_{\psi K_{S}}$ and $\Delta M_{d}/\Delta M_{s}$ are very small
so that these observables determine the coupling $V_{td}$, its phase $-\beta$ and its
magnitude $|V_{td}|$, without significant NP pollution. Therefore, using $S_{\psi K_{S}}$
and $\Delta M_{d}/\Delta M_{s}$ we can construct the Unitary Triangle that is characterized
by $\beta=21.4^{\circ}\pm 1^\circ$ (from $\sin 2\beta\!=\!0.680 \pm 0.025$~\cite{hfag}),
$\gamma=63.5^{\circ}\pm 4.7^\circ$ and $|V_{ub}|\!=\! (3.5\pm 0.2)\cdot 10^{-3}$.

iv) $|\epsilon_K|$ turns out to be uniquely enhanced over its SM value up to a level of
$< 15\%$. This is welcome in view of the decreased value of the parameter
$\hat{B}_{K}$ and the effect is sufficient to reproduce the experimental value of
$|\epsilon_K|$ using the experimental value of $(\sin 2\beta)_{\psi K_{S}}$.

v) Only small effects in $S_{\psi\phi}$ which could, however, be still visible through
the semileptonic asymmetry $A^{s}_{SL}$.

vi) A natural explanation of the $\Delta a_{\mu}$ anomaly (under very mild assumptions).

Finally, we have emphasized that the synergy of high energy and low energy experiments would
provide a unique tool to access information (as the reconstruction of the underlying NP theory)
that cannot be obtained from the LHC or the low energy experiments alone.

It will be very exciting to monitor the upcoming LHC results together with the improved
measurements of $\gamma$, $|V_{ub}|$, $S_{\phi K_{S}}$, $S_{\eta^{\prime} K_{S}}$,
$S_{\psi \phi}$, $A^{s}_{SL}$, $A_{CP}(b\to s\gamma)$ and $d_{e,n}$ as well as improved
evaluation of $\hat{B}_{K}$ in order to see whether new sources of flavor violation beyond
the CKM are required to describe the low energy CP violation in the MSSM framework.\\

\textit{Acknowledgments:}
This work has been supported in part by the Cluster of Excellence ``Origin and Structure
of the Universe'' and by the German Bundesministerium f{\"u}r Bildung und Forschung under
contract 05HT6WOA. We thank M.~Blanke for useful discussions.\\

\textit{Note added:}
We thank Dominik Scherer for pointing out the correct expression for the decay amplitude $A_f$~\cite{Hofer:2009xb} in Eq.~(\ref{defbfu}), that enters the calculation of $S_{\phi K_S}$. According to our new definition for the direct CP asymmetry in $b \to s\gamma$, Eq.~(\ref{acp_bsgamma}), the SM prediction of that observable shifts from $A_{CP}^{\rm SM}(b\to s\gamma) \simeq 0.5\%$ to $A_{CP}^{\rm SM}(b\to s\gamma) \simeq -0.5\%$. Correspondingly, the points in the upper plots of Figs.~\ref{sphiks_cpbsg} and~\ref{epsk} as well as in the lower plot of Fig.~\ref{sphiks_mstop} have been shifted downwards by $\simeq 1\%$.

We also note that the higher order $\tan\beta$ enhanced gluino contributions to $C_{7,8}$ that have been worked out explicitly in~\cite{Hofer:2009xb} are consistently included in our numerical analysis.


\end{document}